\documentclass[aps,pra,showpacs,eqsecnum,floatfix]{revtex4}
\usepackage{graphicx}
\usepackage{dcolumn}
\usepackage{bm}

\title{}

\begin{document}
\title{Efficiency of feedback process in cavity quantum electrodynamics}
\author{H.T.~Fung and
P.T.~Leung\footnote{Correspondence author, email:
ptleung@phy.cuhk.edu.hk}} \affiliation{Department of Physics and
Institute of Theoretical Physics, The Chinese University of
Hong Kong, \\
Shatin, Hong Kong SAR, China}
\date{\today }

\begin{abstract}
Utilizing the continuous frequency mode quantization scheme, we
study from first principle the efficiency of a feedback scheme
that can generate maximally entangled states of two atoms in an
optical cavity through their interactions with a single input
photon. The spectral function of the photon emitted from the
cavity, which will be used as the input of the next round in the
feedback process, is obtained analytically. We find that the
spectral function of the photon is modified in each round and
deviates from the original one. The efficiency of the feedback
scheme consequently deteriorates gradually after several rounds of
operation.
\end{abstract}

 \maketitle
\section{Introduction}
Entanglement has been considered as a fundamental issue to
demarcate the border between quantum and classical physics since
the earliest day of quantum
theory~\cite{Schrodinger-Entanglement,EPR,Bell} and is presently
an essential ingredient in quantum computation
\cite{Chuang-book,Bo-book}. To achieve the construction of quantum
computers, there has been a surge of interest in the generation of
entangled states of atoms and photons through various implements,
including solid-state
devices~\cite{Bouchiat,Bertoni,Friedman,Lloyd-sci-2}, nuclear
magnetic resonance in liquid
samples~\cite{Chuang-sci,Jones-nature}, ion
traps~\cite{Monroe-gate,Turchette-entangle-trap,Sackett-nature},
and atom-photon interactions in optical (or microwave) cavities
~\cite{Kimble-cesium,Haroche-phase-gate,Memory-photon}.

Being a well-studied topic in quantum optics, cavity quantum
electrodynamics (CQED) provides a full arsenal of techniques to
generate entangled atom
pairs~\cite{Braun,Plenio,Duan-multiatom-entangle,Chen,Can} and
also entangled photon
pairs~\cite{Yamamoto-nature,two-photon-imaging}. In particular,
previous studies have shown that cavity loss as well as
environmental noise can help to generate entanglement and increase
the success rate \cite{Plenio,noise-assisted,Chen,Can}. However,
most of these schemes referred above are probabilistic. Therefore,
it presents a challenging task for researchers in this field to
work out a deterministic entangling scheme. With this end in view,
an appealing scheme to generate maximally entangled state of two
atoms inside a leaky cavity in a ``quasi-deterministic" manner has
been proposed \cite{Chen}.  Two $\Lambda$-type three-level atoms,
situated inside a leaky optical cavity and initially prepared in
the ground state $|L\rangle$, interact with a single
left-circularly polarized input photon. The polarization of the
photon emitted from the cavity after the interaction is then
measured. The atoms become maximally entangled if the emitted
photon is right-circularly polarized, or otherwise remain intact.
The probability of success of the entangling process is shown to
approach unity if the input photon is quasi-monochromatic
\cite{Chen}. It is worthwhile to remark that other fundamental
processes in quantum computing like quantum state-swapping and
controlled phase-flip gates can also be achieved with high success
rates via CQED interaction between atoms and such
quasi-monochromatic single photons \cite{swap,cpg}.

Needless to say, the feasibility of the scheme proposed in
\cite{Chen} hinges on the availability of quasi-monochromatic
single photon sources, which is  an issue of current interest in
its own right. On the other hand, it is also possible to
repeatedly redirect the emitted photon into the cavity till it
becomes right-circularly polarized. Hence, the maximally entangled
state can be obtained in a quasi-deterministic way with a single
photon through this feedback process \cite{Hong}. The major
objective of the present paper is to study the efficiency of such
a feedback process, which is crucial to its viability as, due to
effects of absorption and leakage, the process will  inevitably be
terminated after a few rounds. In Ref.~\cite{Hong}  the master
equation approach is adopted. Under the assumption that the
probability of generation of entanglement in each round of the
whole feedback process is a constant value $p$ ($p=1/2$ under the
optimal condition), the overall success rate after $N$ rounds is
given by $1-(1-p)^N$ and hence quasi-deterministic generation of
entanglement is then achievable in the limit of $N \rightarrow
\infty$. However, this crucial assumption has not  been thoroughly
examined in the paper.

The major concern of the master equation approach used in
\cite{Hong} is the evolution of atoms in the presence of cavity
photons inside a leaky cavity. However, the quantum states of the
incident and emitted photons are not properly considered. To
provide a correct description for the incident and emitted
photons, in the present paper we study the same problem using
continuous frequency modes
\cite{Loudon_contd,Lai-narrow-resonance,Scully-MOU,JC-pure-CK} and
describe the quantum states of photons in terms of their spectral
functions \cite{Chen}. In the continuous frequency mode approach
~\cite{Loudon_contd,Lai-narrow-resonance,Scully-MOU,JC-pure-CK},
the incident (emitted) photon is described by a spectral function
$f_{\rm in}(k)$ ($f_{\rm out}(k)$), with $k$ being the wave
number. We show that   if one starts with a cavity photon, which
has a spectral function $f_{\rm c}(k)$ in the form of a complex
Lorentzian with a width given by the leakage rate of the cavity
\cite{Lang_PRA}, the probability of entanglement is $1/2$ under
the optimal condition, agreeing with the result obtained in
\cite{Hong}. However, owing to the interaction with the atom,
$f_{\rm in}(k)$ and $f_{\rm out}(k)$ are different. The spectral
function of the emitted photon is therefore no longer given by
$f_{\rm c}(k)$ despite that the incident photon is a cavity
photon. When the outcome of the measurement is negative, the
emitted photon will be redirected into the cavity. However, as the
initial state of the photon is now different, the operation cannot
be repeated with a constant success rate. Instead, a more
elaborated formulation for the overall success rate has to be
sought.

Using the continuous frequency mode approach, we show in the
present paper
  that the success rate in each round of  the feedback
process decreases gradually. Although the probability of
entanglement does approach unity in the limit of $N \rightarrow
\infty$, the rate at which it approaches unity is much slower that
predicted in Ref.~\cite{Hong}. Therefore, the effects of other
competing loss mechanisms could become crucial in a prolonged
feedback process and great caution has to be exercised in applying
the feedback scheme to ensure quasi-deterministic entanglement
generation.

The structure of the present paper is as follows. In Sect~II we
introduce the system considered in the feedback process
\cite{Chen,Hong} and expand the interaction Hamiltonian in terms
of the continuous frequency
modes~\cite{Loudon_contd,Lai-narrow-resonance,Scully-MOU,JC-pure-CK}.
In Sect.~III we show that the dynamics of the two $\Lambda$-atoms
is reducible to that of a single $\Lambda$-atom and in turn obtain
the spectral function of the emitted photon. We then analyze the
efficiency of the feedback mechanism with our formalism in
Sect.~IV. In Sect.~V we present our conclusion and discuss the
physical significance of the spectral dependence of CQED
processes.

\section{Physical system}
\begin{figure}
\includegraphics[angle=0,width=10.cm]{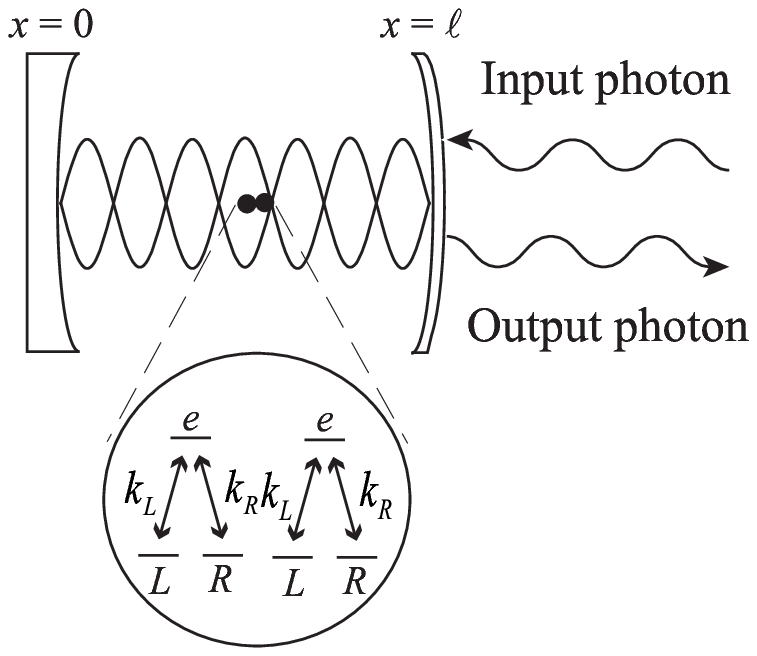}
\caption{Two $\Lambda$-atoms in a leaky cavity.} \label{system2}
\end{figure}

The experimental setup considered in Ref.~\cite{Hong} is
schematically sketched  in Fig.~\ref{system2}. Two identical
$\Lambda$-atoms, $A$ and $B$, are placed at the center of
 a one-dimensional one-sided leaky optical
cavity, which has a  length $l$ and is bounded by two mirrors. The
left mirror at $x=0$ is perfectly reflecting while the right
mirror at $x=l$ is partially transparent.

The normal modes of this leaky cavity are characterized by a
continuous wave number $k$, and are given by
~\cite{Loudon,Chen,Lang_PRA}
\begin{equation}
\label{Uk}
U_k(x)=\left\{\begin{array}{c} I(k)\sin{kx}\\
e^{-ikx}+R(k)e^{ikx} \end{array} \right.
\begin{array}{ccc}{}&{}&{}\\
{}&{}&{}
\end{array}
\begin{array}{c}
0<x\leq l,\\
l<x<\infty,
\end{array}
\end{equation}
where
\begin{eqnarray}
\label{Ik}
I(k)&=&\frac{-2it}{1+re^{2ikl}}\,, \\
\label{Rk} R(k) &=& \frac{-r-t+re^{-2ikl}}{1+re^{2ikl}}\,,
\label{nmode}
\end{eqnarray}
with $r$ and $t$ being the reflection and the transmission
coefficients of the right mirror, respectively. For a good optical
cavity whose transmission coefficient $|t| \ll 1$, $|I(k)|$
sharply peaks at the quasi-mode frequencies
\cite{Loudon,Chen,Lang_PRA} and is negligibly small otherwise.
Besides,  if $k$ is close to a quasi-mode frequency $k_c$, $I(k)$
can be approximated by a single complex lorentzian such that
\begin{equation}\label{lorentz}
|I(k)| \simeq \sqrt{\frac{\kappa}
{2\pi}}\frac{1}{|k-k_c+i\kappa/2|}\;,
\end{equation}
and the width of the lorentzian, $\kappa$, is a measure of the
inverse of the lifetime of the quasi-mode. In fact,
Eq.~(\ref{lorentz}) is a generic result that holds for any
cavities with small leakage rates.

The atomic ground states $|L\rangle$ and $|R\rangle$ are
degenerate and separated from the excited state $|e\rangle$ by an
excitation energy $\omega_e$ . Each atom in its $|L\rangle$
($|R\rangle$) ground state can be excited to $|e\rangle$ by
absorbing a photon in the left-polarized state $|k_L\rangle$
(right-polarized state $|k_R\rangle$), where $k_L$ ($k_R$)
signifies the momentum of the photon. As is assumed in
Ref.~\cite{Hong}, the separation between the atoms is
comparatively small and therefore the atoms have almost the same
coupling strength with the photon field. The Hamiltonian of this
system, in units of $\hbar=c=1$, is given by:
\begin{eqnarray}
\hat{H}=\sum\limits_{\alpha=A,B}\omega_{e}|e_{\alpha}\rangle
\langle e_{\alpha }|+\int_0^{\infty} dk~k
\sum\limits_{\mu=L,R}a_{k\mu
}^{\dag }a_{k\mu }, \\
+\sum\limits_{\alpha=A,B \atop \mu=L,R}\int_0^{\infty} dk \,
a_{k\mu} g_{\mu}(k)|e_{\alpha}\rangle \langle
\mu_{\alpha}|+\mathrm{h}.\mathrm{c}. .
\end{eqnarray}
Here the subscripts $A$ and $B$ respectively label relevant
quantum states of atoms $A$ and $B$. $\hat{a}_{k\mu}$ and
$\hat{a}^{\dagger}_{k\mu}$ are, respectively, the annihilation and
creation operators of a photon with frequency $k$ and polarization
$\mu=L,R$. $g_{\mu}(k)$ is the  dipole coupling strength, which is
frequency-dependent and proportional to $I(k)$. Therefore, under
the premise that the excitation energy $\omega_e$ is close to a
quasi-mode frequency $k_c$, which is assumed in Ref.~\cite{Hong},
$g_{\mu}(k)$ is approximately given by
\begin{equation}
g_{\mu}(k)=\sqrt{\frac{\kappa}{2\pi}}\frac{\lambda_{\mu}}
{k-k_c+i\kappa/2}.
\end{equation}
Here
\begin{equation}
\lambda_{\mu}=\left[\int_{-\infty}^{\infty}dk~|g_{\mu}(k)|^2\right]^{1/2}\;
\end{equation}
is a measure of the dipole moment of the relevant atomic
transition.

As considered in Ref.~\cite{Hong}, initially the two
$\Lambda$-type atoms are prepared in the ground state $|L\rangle$,
while the photon is left-polarized with a normalized spectral
function $f(k')$:
\begin{equation}
|\psi\rangle_{\rm in}=\int^\infty_{-\infty} dk'
f(k')|LL;k'_L\rangle \,,
\end{equation}
where, as usual, we have extended the lower limit of the
integration from $0$ to $-\infty$ and
\begin{equation}
\int_{-\infty}^{\infty}dk'~|f(k')|^2=1.
\end{equation}

\section{Photon-atom interactions}
The quantum dynamics of our system can be solved analytically by
introducing a new set of bases of the Hilbert space \cite{Chen}:
\begin{eqnarray}
|E\rangle&=&\frac{1}{\sqrt{2}}(|eL;0\rangle+|Le;0\rangle),\\
|\Phi\rangle&=&\frac{1}{\sqrt{2}}(|LR\rangle+|RL\rangle).
\label{tranbase}
\end{eqnarray}
Hereafter $|eL\rangle$ denotes product state of $|e_A\rangle$ and
$|L_B\rangle$ and analogous notations will be used in our paper.
As a result of the initial condition considered in the experiment,
an effective Hamiltonian that can fully describe the dynamics of
the system is obtained:
\begin{eqnarray}
\hat{H}_{\rm eff}=&&\!\!\!\!\!\omega_e|E\rangle\langle
E|+\left[\int^{\infty}_{-\infty}\! dk~ k(|LL;k_L\rangle\langle
LL;k_L|+|\Phi;k_R\rangle\langle \Phi;k_R|) +\mathrm{h}.\mathrm{c}.
\right] \nonumber\\&&\!\!\!\!\! +\left[\int^{\infty}_{-\infty}dk~
(\sqrt{2}g_L(k)|E\rangle\langle LL;k_L| +g_R(k)|E\rangle\langle
\Phi;k_R|)\nonumber +\mathrm{h}.\mathrm{c}. \right].
\label{twoatomHam}
\end{eqnarray}

The effective Hamiltonian $\hat{H}_{\rm eff}$ can be furthered
simplified by introducing the states \cite{Chen}:
\begin{eqnarray}
\label{psi1} |\psi_1(k)\rangle &=&
\frac{1}{V(k)}\left(\sqrt{2}g_L^{\ast}(k)|LL;k_L\rangle+g_R^{\ast}(k)|
\Phi;k_R\rangle\right)
, \\
\label{psi2} |\psi_2(k)\rangle &=&
\frac{1}{V(k)}\left(g_R(k)|LL;k_L\rangle-\sqrt{2}g_L(k)|\Phi;k_R\rangle\right)
,
\end{eqnarray}
where
\begin{equation}
\label{Vk} V(k)=\sqrt{2|g_L(k)|^2+|g_R(k)|^2}.
\end{equation}
In terms of these states $\hat{H}_{\rm eff}$ can thus  be written
as \cite{Chen}:
\begin{equation}
\hat{H}_{\rm eff}=\hat{H}_0+\hat{V}+\hat{H}_{\rm dark},
\end{equation}
where
\begin{eqnarray}
\label{H0-eff-psi} \hat{H}_{0}&=&\omega_{e}|E\rangle \langle
E|+\int_{-\infty}^{\infty} dk \, k
|\psi_1(k)\rangle\langle\psi_1(k)|, \\
\label{V-eff-psi} \hat{V}&=&\int_{-\infty}^{\infty} dk \,
V(k)|E\rangle\langle\psi_1(k)|+\mathrm{h}.\mathrm{c}., \\
\label{Hdark} \hat{H}_{\rm dark}&=&\int_{-\infty}^{\infty}dk \, k
|\psi_2(k)\rangle\langle\psi_2(k)|,
\end{eqnarray}
with $\hat{H}_{\rm dark}$ being the free Hamiltonian of the dark
states $|\psi_2(k)\rangle$ that does not take part in the
interaction. Therefore, the system is now analogous to a two-level
atom and we will omit the term $\hat{H}_{\rm dark}$ in the
following discussion.

The evolution of the system at a later time $t>0$ can be obtained
once the retarded Green's function,
$\theta(t)\exp[-i(\hat{H}_0+\hat{V}) t]$, is known. To this end,
we consider the resolvent of the Hamiltonian $\hat{H}_0+\hat{V}$,
which is defined by \cite{Cohen}:
\begin{equation}
\hat{G}(\omega)=\frac{1}{\omega-\hat{H}_0-\hat{V}} ,  \label{Gw}
\end{equation}
with $\omega$ being a complex variable. It yields the retarded
Green's function, $\theta(t)\exp[-i(\hat{H}_0+\hat{V}) t]$,
through an integral transformation:
\begin{equation}
\theta(t)\exp[-i(\hat{H}_0+\hat{V}) t]
=\lim_{\epsilon\rightarrow0^+}\frac{i}{2\pi}\int_{-\infty+i\epsilon}
^{\infty+i\epsilon}
\hat{G}(\omega)e^{-i\omega t}d\omega. \label{Evo-Gw}
\end{equation}
The exact form of the resolvent $\hat{G}(\omega)$ has been worked
out in Ref.~\cite{Chen}. In the following we shall make use of it
to evaluate the final state of our system subject to two different
initial conditions.

As considered in Ref.~\cite{Hong}, the two atoms first interact
with an photon existing in the cavity. Such a photon is termed as
a cavity photon (or a quasi-mode photon) \cite{Plank}, which is
characterized by a spectral function $f(k')=f_{\rm c}(k')$ given
by:
\begin{eqnarray}
f_{\rm c}(k')\equiv\frac{1}{\lambda_{\mu}}g_{\mu}^*(k')=
\sqrt{\frac{\kappa} {2\pi}}\frac{1}{k'-k_c+i\kappa/2}.
\label{cavityform}
\end{eqnarray}
At $t=0$ the cavity photon is confined in the cavity and leaks out
of the cavity at a rate $\kappa$ for $t>0$. If a left-polarized
cavity photon is prepared in the cavity at $t=0$, the initial
state can be represented as:
\begin{equation}
|\psi\rangle_{\rm in}=\int_{-\infty}^{\infty}
dk^{\prime}~\frac{f_{\rm c}(k^{\prime})}{V(k^{\prime})}\left(
\sqrt{2}g_L(k^{\prime})|\psi_1(k^{\prime})\rangle
+g_R^{\ast}(k^{\prime})|\psi_2(k^{\prime}\rangle)\right) .
\end{equation}
By using the resolvent method developed in Ref.~\cite{Chen}, we
obtain the output state of the first round:
\begin{equation}
|\psi \rangle _{{\rm out},1} =|LL\rangle \otimes
\int_{-\infty}^{\infty} dk ~ f_{\rm c}(k)
D_L(k)e^{-ikt}|k_{L}\rangle +|\Phi \rangle \otimes
\int_{-\infty}^{\infty} dk~ f_{\rm c}(k)
 D_R(k) e^{-ikt}|k_{R}\rangle , \label{cav1}
\end{equation}
where
\begin{eqnarray}
D_L(k)&=&\frac{(\Delta k-\delta_e)(\Delta
k+i\kappa/2)-\lambda_R^2} {(\Delta k-\omega_+)(\Delta k-\omega_-)}
\label{CL}
\,,\\
D_R(k)&=&\frac{\sqrt{2}\lambda_{L}\lambda_{R}} {(\Delta
k-\omega_+)(\Delta k-\omega_-)}\,, \label{CR}
\end{eqnarray}
and the complex Rabi frequencies $\omega_{\pm}$ are defined by:
\begin{equation}
\omega_{\pm}=\frac{\delta_{e}-i\kappa/2}{2}\pm\sqrt{(\frac{\delta_{e}
+i\kappa/2}{2})^2+2\lambda_{L}^2+\lambda_{R}^2}\,,
\end{equation}
with the detuning $\delta_{e}=\omega_e-k_c$. Since the output
state $|\psi \rangle _{\rm out}$ is also normalized, it is obvious
that:
\begin{equation}
\int_{-\infty}^{\infty} dk ~ |f_{\rm c}(k) D_L(k)|^2 +
\int_{-\infty}^{\infty} dk~ |f_{\rm c}(k)
 D_R(k)|^2 =1 \,. \label{cav2}
\end{equation}

After the first encounter with the atoms, the photon is emitted
from the cavity and its polarization is detected. If the emitted
photon is right-polarized, then the two atoms will be maximally
entangled as indicated in Eq.~(\ref{cav1}). On the other hand, if
the emitted photon is left-polarized, the two atoms will remain in
the state $|LL\rangle$ and the emitted photon will be sent back to
the cavity. In this case, the two atoms have to interact with a
photon injected from the exterior of the cavity. Such an initial
photon satisfies the scattering condition \cite{Chen}, and  has a
modified normalized spectral function $f_n(k)$ after $n$ rounds of
interactions. As a result, the final state of our system after the
$(n+1)$-th interaction is then given by \cite{Chen}:
\begin{eqnarray}
|\psi\rangle_{{\rm
out},n+1}=|LL\rangle\otimes\int_{-\infty}^{\infty}
dk~f_n(k)C_{L}(k) e^{-ikt}|k_L\rangle\nonumber\\
-|\Phi\rangle\otimes\int_ {-\infty}^{\infty}dk~f_n(k)C_{R}(k)
e^{-ikt}|k_R\rangle, \label{scat1}
\end{eqnarray}
where
\begin{widetext}
\begin{eqnarray}
C_{L}(k)&=&\frac{(\Delta k-\delta_{e})(\Delta
k^2+\kappa^2/4)-\Delta
k(\lambda_{R}^2+2\lambda_{L}^2)+i\kappa(\lambda_{R}^2-2\lambda_{L}^2)/2}
{(\Delta k-i\kappa/2)(\Delta k -\omega_{+})(\Delta k-\omega_{-})},\\
C_{R}(k)&=&\frac{\sqrt{2}i\kappa\lambda_{L}\lambda_{R}} {(\Delta
k-i\kappa/2)(\Delta k-\omega_{+})(\Delta k -\omega_{-})}.
\label{scat1R}
\end{eqnarray}
\end{widetext}
It is interesting to note that $C_R(k)$ and $C_L(k)$ themselves
satisfy the unitarity condition:
\begin{equation}
|C_L(k)|^2+|C_R(k)|^2=1\;, \label{scat2}
\end{equation}
reflecting the fact that waves with different frequencies are not
mixed upon such scattering.
\section{Feedback mechanism}
From the functions $D_L(k)$, $D_R(k)$, $C_L(k)$ and $C_R(k)$
obtained in the previous section, one can analyze the feedback
mechanism initiated by an initial cavity photon \cite{Hong}. After
the first round, the output state can be obtained from
Eq.~(\ref{cav1}). The probability of getting a right-polarized
photon and hence a maximally entangled state in the first round is
given by:
\begin{equation}\label{p1}
p_{1}^R=\int^{\infty}_{-\infty} dk|D_R(k)f_{\rm c}(k)|^2
\end{equation}
Under the optimal condition where $\lambda_R=\sqrt{2}\lambda_L$
and in the strong coupling limit, $p_{1}^R$ is equal to $1/2$
\cite{Chen}. This is in perfect agreement with the numerical
result obtained in Ref.~\cite{Hong}.

However, if the emitted photon is left-polarized, which has a
normalized spectral function:
\begin{equation}
f_1(k)=\frac{D_{L}(k)f_{\rm c}(k)}{\left[\int_{-\infty}^{\infty}dk
|D_L(k)f_{\rm c}(k)|^2\right]^{1/2}}, \label{sf1}
\end{equation}
it will be re-injected into the cavity as the initial state of the
second round. As $f_1(k)$ is in general different from $f_{\rm
c}(k)$, its interaction with the two atoms will not be the same as
that of the initial cavity photon. Therefore, the efficiency of
the feedback process has to be examined with extra care.

Disregarding the difference in the spectral functions of the
incident photons, the dynamics of the second round is exactly the
same as the first one and the output state is given by
Eq.~(\ref{scat1}). The probability of getting a left-polarized
photon in the first round and a right-polarized photon in the
second round is:
\begin{eqnarray}
p^R_{2}&=&\int^{\infty}_{-\infty} dk |C_R(k) f_1(k)|^2 \times
\int^{\infty}_{-\infty} dk |D_L(k)f_{\rm c}(k)|^2 \nonumber \\
&=&\int^{\infty}_{-\infty} dk | C_R(k)D_L(k)f_{\rm c}(k)|^2
\end{eqnarray}

The photon leaks out from the cavity after its second encounter
with the atom and its polarization is measured. If once again, a
left-polarized photon is detected, the normalized spectral
function of this photon is given by:
\begin{eqnarray}
f_2(k)&=&\frac{C_L(k)f_1(k)}
{\left[\int_{-\infty}^{\infty}dk|C_L(k)f_1(k)|^2\right]^{1/2}}
\nonumber
\\ &=&\frac{C_L(k)D_L(k)f_{\rm c}(k)}
{\left[\int_{-\infty}^{\infty}dk|C_L(k)D_L(k)f_{\rm
c}(k)|^2\right]^{1/2}}.
\end{eqnarray}
 This feedback process is
carried on iteratively until a right-polarized photon is detected
outside the cavity. After $n$ rounds of interactions, the
normalized spectral function of the left-polarized photon can be
found by using Eqs.~(\ref{cav1}) and (\ref{scat1}), yielding
\begin{eqnarray}
f_{n}(k)&=&\frac{C_L(k)f_{n-1}(k)}
{\left[\int_{-\infty}^{\infty}dk|C_L(k)f_{n-1}(k)|^2\right]^{1/2}}
\nonumber \\ &=&\frac{[C_L(k)]^{n-1}D_L(k)f_{\rm c}(k)}
{\left[\int_{-\infty}^{\infty}dk|[C_L(k)]^{n-1}D_L(k)f_{\rm
c}(k)|^2\right]^{1/2}}.
\end{eqnarray}

Following the argument developed above and using mathematical
induction, one can readily show that the probability of getting a
left-polarized photon in the first $(n-1)$ rounds and a
right-polarized one in the $n$-th round ($n\geq2$) is given by:
\begin{eqnarray}
p^R_{n}&=&p^L_{n-1}\int^{\infty}_{-\infty} dk \, |
C_R(k)f_{n-1}(k)|^2 \nonumber \\
&=&\int^{\infty}_{-\infty} dk \, |
C_R(k)[C_L(k)]^{n-2}D_L(k)f_{\rm c}(k)|^2,
\end{eqnarray}
where $p^L_{n-1}$ is the probability of detecting a left-polarized
photon in the first $n-1$ round and is given by:
\begin{equation}
p^L_{n-1}=\int^{\infty}_{-\infty}
dk\,|[C_{L}(k)]^{n-2}D_{L}(k)f_{\rm c}(k)|^2\,.
\end{equation}
The cumulative probability of generating a pair of maximally
entangled atoms after the first $N$ rounds of interaction is
therefore
\begin{eqnarray}
P^R_{N}&=&\sum_{n=1}^Np^R_n\\
&=&1-\int^{\infty}_{-\infty} dk | [C_{L}(k)]^{N-1}D_{L}(k)f_{\rm
c}(k)|^2\,,
\end{eqnarray}
where the unitarity condition (\ref{scat2}) has been applied to
simplify the expression. Since the magnitude of $C_{L}(k)$ is
always less than unity, it is obvious that
\begin{equation}
\lim_{N\to\infty}P^R_{N}=1, \label{probr}
\end{equation}
leading to the conclusion that after a sufficiently large number
of rounds the two atoms will surely become maximally entangled.
However, the limit $N\to\infty$ in Eq.~(\ref{probr}) can be taken
only if other sources of energy leakage can be safely ignored. To
eliminate (or at least minimize) the effects arising from photon
loss and atomic spontaneous decay, it is advantageous to get a
high enough success probability in a few rounds. Otherwise, the
plausibility of the feedback mechanism to ensure
quasi-deterministic entanglement still remains precarious.
Therefore, a study of rate at which $P^R_{N}$ approaches unity is
called for.

\begin{figure}
\includegraphics[angle=0,width=10.cm]{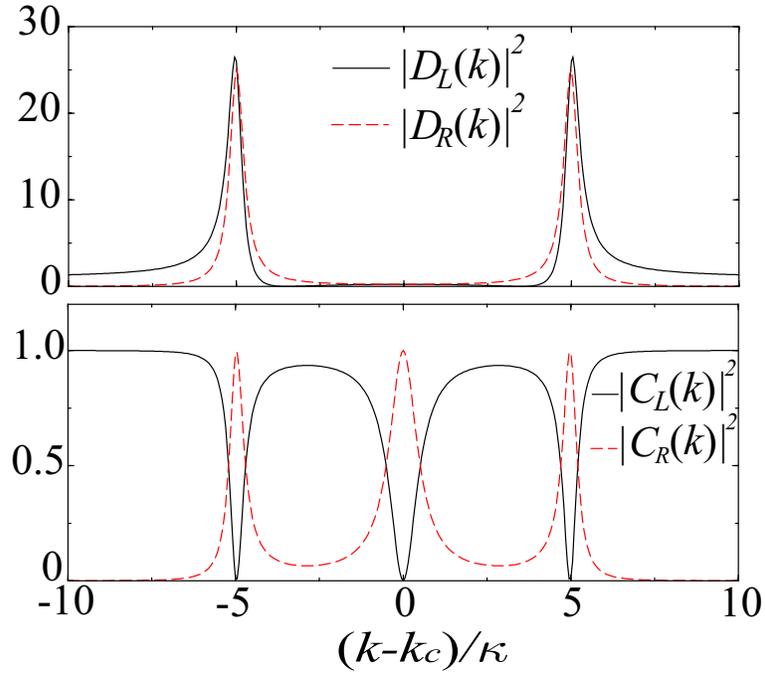}
\caption{(Color online) (a) $|D_L(k)|^2$ (solid line),
$|D_R(k)|^2$ (dashed line); and (b) $|C_L(k)|^2$ (solid line),
$|C_R(k)|^2$ (dashed line) are plotted for the optimal case
($\lambda_R=\sqrt{2}\lambda_L$) with $\lambda_L=2.5\kappa$.}
\label{CLCR}
\end{figure}

\begin{figure}
\includegraphics[angle=0,width=10.cm]{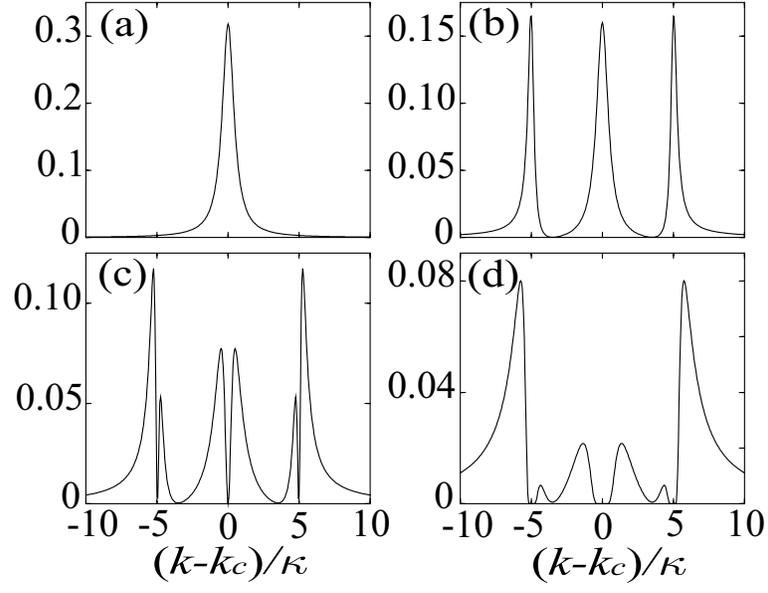}
\caption{A plot of (a) $|f_{\rm c}(k)|^2$, (b) $|f_1(k)|^2$, (c)
$|f_2(k)|^2$ and (d) $|f_{10}(k)|^2$ for the optimal case
($\lambda_R=\sqrt{2}\lambda_L$) with $\lambda_L=2.5\kappa$}
\label{all}
\end{figure}

\begin{figure}
\includegraphics[angle=0,width=10.cm]{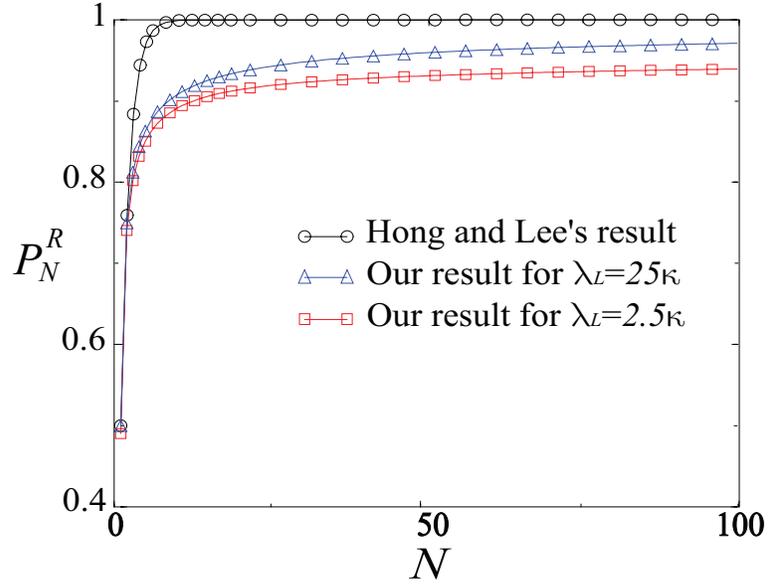}
\caption{(Color online) Probability of  entanglement generation
within $N$ rounds of operation for the optimal case
($\lambda_R=\sqrt{2}\lambda_L$). The triangles
($\lambda_L=25\kappa$) and the squares ($\lambda_L=2.5\kappa$)
show the results obtained from our theory, while the circles are
data obtained from the constant success rate assumption with
$p=1/2$. } \label{onepn}
\end{figure}

In Fig.~\ref{CLCR} we show the absolute values of $C_{L(R)}$ and
$D_{L(R)}$ for an optimal case with $\delta_e=0$, $\kappa=1$ and
$\lambda_L=\lambda_R/\sqrt{2}=2.5\kappa  $ \cite{Hong,Chen}.
Figure~\ref{all} clearly demonstrates the variation in the
spectral function of the photon during the feedback process.
Contrary to the behavior of $|f_1(k)|$, which have three maxima at
$\Delta k \equiv k-k_c=0,\omega_{\pm}\,$, $|C_L(k)|$ vanishes at
$\Delta k =0,\omega_{\pm}$ and is quite small around there.
$p^R_n$ consequently decreases significantly as $n$ increases. As
shown in Fig.~\ref{onepn}, the total probability of entanglement
after $N$ rounds of operation is almost saturated after about 10
rounds and is still away from unity even after 100 rounds. This
differs markedly from the result predicted in Ref.~\cite{Hong},
which converges to unity rapidly. We also note that the saturation
phenomenon persists for another case with
$\lambda_L=\lambda_R/\sqrt{2}=25\kappa$. Therefore, the efficiency
of the feedback scheme remains low even for a much larger
interaction strength.

\section{Conclusion}
Adopting the continuous frequency mode quantization scheme, we
studied the evolution of a photon under the feedback mechanism and
showed that the spectral function of the photon is in general
modified in each round of the feedback process. Thus, the feedback
scheme is not as effective as what Ref.~\cite{Hong} has claimed.
Whether the above-mentioned scheme can ensure quasi-deterministic
entanglement after a large number of iterations depends crucially
on the possibility of elimination of other competing processes,
say, photon losses in the cavity and in the optical fibres.
Nevertheless, the feedback scheme could still be a useful method
to improve the probability of success until other mechanisms might
interfere with its operation.

The aim of the feedback process proposed in Ref.~\cite{Hong}  is
to ensure quasi-deterministic generation of entangled states with
only one single input photon. If, instead, multiple independent
cavity photons are used, the probability of success in each
interaction is at best 0.5 \cite{Hong}. Besides, as argued by Hong
and Lee in Ref.~\cite{Hong}, for an experiment using multiple
photons the finite detection efficiency of the detector $D_2$ that
measures right-polarized output photons may give rise to ambiguity
stated as follows. When no photon is detected by the detector
$D_2$, there are two possible cases: (1) the entanglement process
has not yet succeeded, or (2) the detector $D_2$ failed to detect
the output photon. The introduction of the feedback process, which
has a success rate close to unity, in effect makes the detector
$D_2$ redundant and eliminates the ambiguity mentioned above.

On the other hand, it is worthwhile to note the accumulative
effect resulting from the difference between the spectra of the
input and output photons. As demonstrated in the present paper
(see Fig.~3), the spectrum of an input photon is in general
different from that of the output photon. This point is often
overlooked in the conventional master equation approach to leaky
cavities. If the output photon is then used as the input of a new
 process, such difference in the spectra might give rise to
accumulative effect and harm the overall performance. Needless to
say, photons are ideal mediators of quantum information and can
generate entanglement between well separated atoms. However, one
must be conscious of the change in the spectra of a photon before
and after interacting with an atom as such change might lead to
marked difference in the long run.

Finally, we would also like to remark that the probability of
success in the process examined here depends sensitively on the
spectrum of the input photon. For example, instead of using cavity
photons with a spectral width equal to that of the cavity, one can
also consider an input photon with an arbitrary width $\kappa_{\rm
in}$, i.e.
\begin{eqnarray}
f(k')= \sqrt{\frac{\kappa_{\rm in}}
{2\pi}}\frac{1}{k'-k_c+i\kappa_{\rm in}/2}. \label{cavityform}
\end{eqnarray}
As shown in Fig.~5, for a ``quasi-monochromatic" photon which has
a sufficiently narrow spectral width, the probability of success
in one single trial can be close to unity irrespective of the
value of $\lambda_L/\kappa$ \cite{Chen}. Similarly,
\cite{swap,cpg} suggest that quantum state-swapping and controlled
phase-flip gates are achievable with high success rates with
quasi-monochromatic single photons. Therefore, the quantum state
of single photons is likely to play a crucial role in quantum
computing.
\begin{acknowledgments}
We thank TW~Chen for helpful discussions and also for his
contribution to the initial stage of this project. The work
reported here is partially supported by two grants from the
Research Grants Council of the Hong Kong SAR, China (Project Nos.
428200 and 401603).
\end{acknowledgments}
%




\newpage


\begin{figure}
\includegraphics[width=8.6cm]{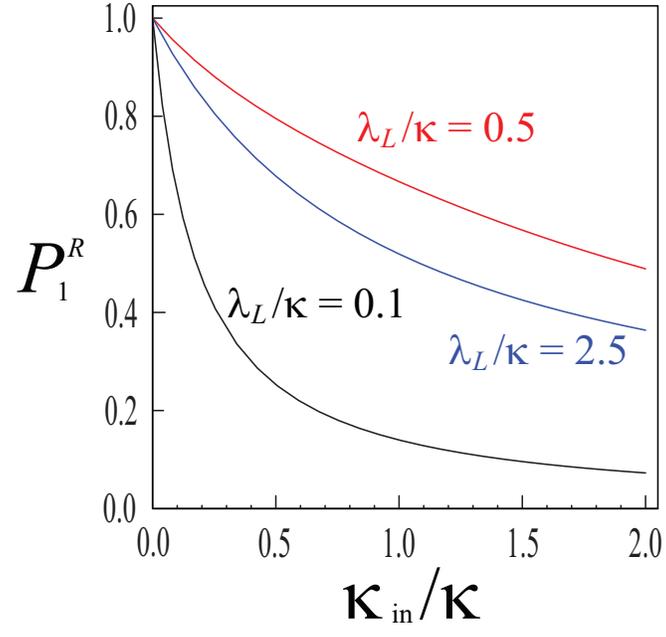}
\caption{(Color online) Probability of  generating the entangled
state in one trial, $P^R_1$, versus $\kappa_{\rm in}/\kappa$ for
an input photon with a Lorentzian spectral function given by
Eq.~(\ref{cavityform}), with $\lambda_R=\sqrt{2}\lambda_L$,
$\lambda_L/\kappa=0.1$, $0.5$, and $2.5$.} \label{fig5}
\end{figure}
\end{document}